\documentclass[twocolumn,secnumarabic,amssymb, amsmath, nofootinbib,tightenlines,
nobibnotes, aps, prl,epsfig]{revtex4}
\usepackage{graphicx}
\usepackage{dcolumn}
\usepackage{bm}
\begin{document}
\preprint{APS/123-QED}
\title{The improved saturation model in nuclei}
\author{G.R.Boroun}%
 \email{boroun@razi.ac.ir }
\author{B.Rezaei}%
 \email{brezaei@razi.ac.ir }

\affiliation{Department of  Physics , Razi University, Kermanshah
67149, Iran}
\date{\today}
\begin{abstract}
We consider the nuclear shadowing in deep-inelastic scattering
corresponding to kinematic regions accessible by future
experiments at electron-ion colliders. The gluon distribution at
small $x$ is obtained using an improved dipole model depended  on
the impact parameter for atomic nucleus and compared with nCETQ15
parametrization group. The nuclear shadowing at small $x$ is
defined within the color dipole formalism with respect to the mass
number $A$. Its behavior is predicted for light nuclei in a wide
range of the impact parameter $b$ and the transverse dipole size
$r$. The nuclear saturation at large-$r$ (small $\mu^2$) is
observable. The behavior of the nuclear ratio
$\sigma^{A}_{\mathrm{dip}}/\sigma_{0}$ is similar to the
Golec-Biernat-W$\ddot{\mathrm{u}}$sthoff (GBW)
 model in a wide range of  $r$ for light and heavy nuclei at small $x$. \\

\end{abstract}
 \pacs{***}
\maketitle
\subsection{I. Introduction}

The structure of hadrons in electron-ion interaction in terms of
quarks and gluon distribution functions (PDFs), in deep inelastic
scattering (DIS), are interesting in the future circular collider
hadron-electron (FCC-he) and the large hadron electron collider
(LHeC)[1]. Study of nuclear structure and nuclear collisions will
be considered on the electron-Ion collider (EIC) [2,3]. The
shadowing effects will be important at small values of the Bjorken
variable $x$ which is a consequence of multiple scattering at high
energies where  a hadron becomes a dense system. Indeed, the QCD
dynamics of the saturation effects will be  visible at small $x$.
This effect is due to the growth of the gluon density with energy
and can be studied in the high-density regime of QCD. Its further
growth is expected to slow down due to the non-linear QCD effects
associated with the unitarity corrections [4-9]. The saturation
scale $Q^{2}_{s}(x)$ is characterized on the saturation approach
and depends on energy. This marks the transition between the
linear and saturation regions.\\
The nuclear photoabsorption cross sections at small $x$ lie on a
single curve when plotted against the variable $Q^{2}/Q^{2}_{s,A}$
[10-12], where $A$ is the number of nucleons in a nuclear target
with $Q^{2}_{s,A}{\propto}A^{j}Q^{2}_{s}$ where $j\simeq
\frac{1}{3}$ or $\frac{4}{9}$ (for large nuclei the value $j$
corresponds to $1/3$) in Refs.[4-6,11-16] and
$Q^{2}_{s}{\sim}x^{-\lambda}$ and $\lambda{\simeq}0.3$. Non-linear
effects in nuclei are expected when
$\alpha_{s}T_{A}(b)xg(x){\sim}Q^2$ because they have more gluons
than protons, where $T_{A}(b)$ is the nuclear thickness and $g(x)$
is the gluon density\footnote{ $G(x,Q^2)=xg(x,Q^2)$ where
$G(x,Q^2)$ is the gluon distribution.}. The nuclear shadowing at
the Bjorken variable
$x{\ll}x_{A}=\frac{1}{m_{N}R_{A}}=0.15A^{-1/3}$ becomes important
(where $R_{A}$ is the radius of the target nucleus and $m_{N}$ is
the nucleon mass [13]).\\
The behavior of the gluon density at very small $x$ describes the
exclusive processes in ep and eA collisions and this is important
for connection of the dipole-target amplitude to the integrated
gluon density. The gluon recombination, in a fast moving frame, is
due to the overlap of the gluon clouds of different nucleons. This
makes gluon density in the nucleus with mass number A smaller than
A times that in a free nucleon. This behavior is shown in Ref.[17]
using the $"$brute force$"$ method in the momentum space.\\
The saturation effects at EIC play an important role in the
processes $e+A{\rightarrow}e+X$ in experiments  with the variable
center-of-mass energy within the range
$20<\sqrt{s}<140~\mathrm{GeV}$. The kinematic regions in
experiments at the proposed EIC at the Brookhaven National
Laboratory give a deeper knowledge of  hadronic structure at high
energies [2,3]. This energy is lower than at HERA with
$\sqrt{s}=318~\mathrm{GeV}$ but the luminosity is higher by a
factor of 1000. The EIC will combine the experiences of HERA and
RHIC, which will have a strong impact on understanding the small
and large-$x$ regions of nuclear shadowing and the EMC effect in
comparison with fixed-target kinematics for various
nuclei [18-20].\\
In this paper we consider the nuclear dipole cross sections in the
region of small $x$ ($x{\leq}0.01$) in the improved dipole picture
with respect to the bSat and bCGC models at the kinematical range
that will be probed by the EIC and LHeC. The paper is organized as
follows. In the next section, we present a brief overview of the
formalism needed for the description of the exclusive processes in
ep collisions and discuss the distinct models for the
dipole-proton scattering amplitude employed in our analysis. In
section III, we exhibit the dipole cross section models in $eA$
collisions with respect to the nuclear gluon density. In Section
IV a comparison of the results with available data on $G^{A}/A$
will be shown and the dipole cross sections will be discussed.
Finally, in the last section conclusions  will be outlined.\\

\subsection{II. The Dipole Cross-Section Model for $\gamma^*$-p}

The scattering between  the virtual photon $\gamma^{*}$ and the
proton is seen as the  color dipole. This color dipole picture is
a factorization scheme for  DIS in electron-proton ($ep$) and
lepton-nucleus ($lA$) scattering. The dipole cross section is
factorized into a light-cone wave function by the following form
\begin{eqnarray}
\sigma_{L,T}^{\gamma^{*}p}(x,Q^{2})=\int dz d^{2}\mathbf{r}
|\Psi_{L,T}(\mathbf{r},z,Q^{2})|^{2}\sigma^{p}_{\mathrm{dip}}(\widetilde{x}_{f},\mathbf{r}).
\end{eqnarray}
where the transverse dipole size $r$ and the
 longitudinal momentum fraction $z$ are defined  into the photon
 momentum [21-26]. Here $\Psi_{L,T}$ are defined by  spin averaged light-cone
wave functions\footnote{Where the subscript $L$ and  $T$ refer to
the transverse and longitudinal polarization state of the
exchanged photon.} of the photon, and
$\sigma_{\mathrm{dip}}(\widetilde{x}_{f},r)$ is the dipole
cross-section. The dipole cross-section contains all the
information about the target and the strong interaction physics,
and it is related to the imaginary part of $(q\overline{q})p$
forward scattering amplitude. The Bjorken variable $x$ is modified
by taking into account the active quark mass as it is equivalent
to $\widetilde{x}_{f}{\equiv}x(1+4m_{f}^{2}/Q^{2})$ where $m_{f}$
is
the mass of the quark of flavor $f$.\\
The dipole cross section in the eikonal-like approach was
proposed\footnote{The GBW model was updated in [15] to improve the
large $Q^{2}$ description of
 the proton structure function by a modification of the small $r$ behavior of the dipole cross
section to include evolution of the  gluon distribution.} [27] by
\begin{eqnarray}
\sigma^{p}_{\mathrm{dip}}(\widetilde{x}_{f},r)=\sigma_{0}(1-e^{-r^{2}Q^{2}_{s}/4}).
\end{eqnarray}
 The dipole cross section shows the colour transparency property when
$r{\rightarrow}0$, i.e. $\sigma_{\mathrm{dip}}\sim r^{2}$, which
is pQCD phenomenon and the saturation property at large $r$, i.e.
$\sigma_{\mathrm{dip}}\sim \sigma_{0}$ , which satisfies the
unitarity condition.  The dipole cross section improved by
Bartels-Golec-Biernat-Kowalski (BGBK) [24] with adding the
collinear effects\footnote{Although BGBK model is  successful in
describing dipole cross section at large values of $r$ as the two
models (GBW and BGBK) overlap in this region but they differ in
the small $r$ region where the running of the gluon distribution
starts to play a significant role. Indeed the improved model of
$\sigma_{\mathrm{dip}}$ significantly improves agreement at large
values of $Q^{2}$ without affecting the physics of saturation
responsible for transition to small $Q^{2}$.}. The dipole cross
section with implementation of QCD evolution  on the gluon
distribution reads
\begin{eqnarray}
\sigma^{p}_{\mathrm{dip}}=\sigma_{0}\{1-\exp(-\frac{\pi^{2}r^{2}\alpha_{s}(\mu^{2})xg(\widetilde{x}_{f},\mu^{2})}{3\sigma_{0}})\},
\end{eqnarray}
where the hard scale $\mu$ is  assumed to have the form
\begin{eqnarray}
\mu^{2}=C/r^{2}+\mu^{2}_{0},
\end{eqnarray}
with the parameters $C$ and $\mu^{2}_{0}$ where they are obtained
from a fit to the DIS data [24].\\
By introducing the impact parameter (IP) of the proton, the dipole
cross section reads
\begin{eqnarray}
\sigma^{p}_{\mathrm{dip}}(x,r)=\int
d^{2}b\frac{d\sigma^{p}_{\mathrm{dip}}}{d^{2}b}
\end{eqnarray}
where $b$ is the impact parameter (IP) of the center of the dipole
relative to the center of the proton, and
\begin{eqnarray}
\frac{d\sigma^{p}_{\mathrm{dip}}}{d^{2}b}=2(1-\mathrm{Re}~S(b)),
\end{eqnarray}
where $S(b)$ is the S-matrix element of the elastic scattering,
and it is proportional to the dipole area, the strong coupling,
the number of gluons in the cloud and the shape function as
\begin{eqnarray}
\frac{d\sigma^{p}_{\mathrm{dip}}}{d^{2}b}=2\Big{[}1-
\exp\Big{(}-\frac{\pi^{2}r^{2}\alpha_{s}(\mu^{2})xg(\widetilde{x}_{f},\mu^{2})T(b)}{2N_{c}}\Big{)}
 \Big{]}.
\end{eqnarray}
Here, the function $T(b)$ is determined from a fit to the data by
the exponential form
\begin{eqnarray}
T(b)=\frac{1}{2{\pi}B_{G}}\exp(-b^{2}/2B_{G}),
\end{eqnarray}
where the parameter $B_{G}$ was found to be
$4.25~\mathrm{GeV}^{-2}$ [25].\\
For multi Pomeron exchange
${d\sigma_{\mathrm{dip}}}/{d^{2}b}=2N(x,r,b)$, where the
eikonalised dipole scattering amplitude can be expanded as
\begin{eqnarray}
N(x,r,b)=\sum_{n=1}^{\infty}\frac{(-1)^{n+1}}{n!}
\Big{[}\frac{\pi^{2}}{2N_{c}}r^{2}\alpha_{s}(\mu^{2})xg(\widetilde{x}_{f},\mu^{2})T(b)\Big{]}^{n},\nonumber\\
\end{eqnarray}
 where the $n$-th term in the expansion corresponds to $n$-Pomeron
exchange [25]. In the Color Glass Condensate (CGC) effective
theory [28-29] the dipole cross section at small $r$  (i.e.,
Eq.(8)) reads
\begin{eqnarray}
\frac{d\sigma^{p}_{\mathrm{dip}}}{d^{2}b}=\frac{\pi^{2}r^{2}\alpha_{s}(\mu^{2})xg(\widetilde{x}_{f},\mu^{2})T(b)}{N_{c}}.
\end{eqnarray}
The BGBK  and CGC models with impact parameters are denoted by the
IP-Sat and  b-CGC models respectively\footnote{The
Balitsky-Kovchegov (BK) equation [30-32] for a dipole scattering
amplitude was proposed in terms of the hierarchy of equations for
Wilson line operators in the limit of large number of colors
$N_{c}$. The geometrical scaling (GS) [33] at the high-energy
limit of perturbative QCD is obtained from the BK equation [30-32]
and the CGC formalism [34]. The BGBK  and CGC models considered
only the dipole cross section integrated over the impact parameter
$b$ [35].}. The impact parameter dependence in the b-CGC model of
the saturation scale $Q_{s}$ was introduced [35] by
\begin{eqnarray}
Q_{s}{\equiv}Q_{s}(x,b)=(\frac{x_{0}}{x})^{\lambda/2}\Big{[}\exp(-\frac{b^{2}}
{2B_{CGC}})\Big{]}^{1/2\gamma_{s}},
\end{eqnarray}
where  the parameter $B_{CGC}$ is a free parameter and is
determined by the $t$ distribution of the exclusive
diffractive processes at HERA.\\

\subsection{III. The Dipole Cross-Section Model for $\gamma^*$-A}

The saturation scale in $\gamma^*$-A interactions, $Q_{s,A}^{2}$,
is defined [8] from the running of the coupling by the following
form
\begin{eqnarray}
Q_{s,A}^{2}{\ln}\bigg{(}\frac{Q_{s,A}^{2}}{\Lambda^{2}_{\mathrm{QCD}}}\bigg{)}{\propto}
\bigg{(}\frac{T_{A}(b)}{T_{A}(0)} \bigg{)}
A^{1/3}Q_{s}^{2}{\ln}\bigg{(}\frac{Q_{s}^{2}}{\Lambda^{2}_{\mathrm{QCD}}}\bigg{)},
\end{eqnarray}
where $T_{A}(b)$ is the nuclear profile function normalized to
unity, $\int d^2b~ T_{A}(b)=1$. Here $b$ is the impact parameter
(IP) of the center of the dipole relative to the center of the
nucleus. In the limit $r{\rightarrow}0$, Eq.(12) is rewritten by
applying the first scattering approximation in the dipole cross
section in the form
$Q_{s,A}^{2}=\frac{1}{2}AT_{A}(b)\sigma_{0}Q_{s}^{2}$. In momentum
space, the saturation scale is obtained according to the maximum
of the unintegrated gluon distribution by
\begin{eqnarray}
Q_{s,A}^{2}{\simeq}\bigg{[}4{\ln}\bigg{(}
\frac{2AT_{A}(b)\sigma_{0}}{2AT_{A}(b)\sigma_{0}-1}\bigg{)}
\bigg{]}^{-1}Q_{s}^{2}.
\end{eqnarray}
Here $T_{A}(b)$ is the nuclear thickness function where is defined
in Ref.[36] by
\begin{eqnarray}
T_{A}(b)=\frac{3R_{A}}{2{\pi}r_{0}^{3}}\sqrt{1-\frac{b^2}{R_{A}^2}}.
\end{eqnarray}
This is obtained from a hard-sphere model for nuclear distribution
in the rest frame
\begin{eqnarray}
\rho_{A}(r)=\frac{3}{4{\pi}r_{0}^{2}}\theta(R_{A}-r),
\end{eqnarray}
 as
 \begin{eqnarray}
Q_{s,A}^{2}{\approx}Q_{s}^{2}A^{1/3}\sqrt{1-\frac{b^2}{R^{2}_{A}}},
\end{eqnarray}
with $r_{0}=1.12~\mathrm{fm}$. The nuclear thickness function in
the Woods-Saxon distribution by assuming that the positions of the
nucleons $\{\mathbf{b}_{i}\}$ are distributed reads [7]
\begin{eqnarray}
T_{A}(b)=\int dz\frac{C}{1+\exp[(\sqrt{b^2+z^2}-R_{A})/d]}.
\end{eqnarray}
The Woods-Saxon distribution is used for $A>20$ and for light
nuclei ($A<20$) a gaussian profile is used [37] by the following
form
\begin{eqnarray}
T_{A}(b)=\frac{3}{2{\pi}R^2_{A}}\exp(-3b^2/2R^2_{A}),
\end{eqnarray}
where the nuclear radius parametrized as
$R_{A}=0.82A^{1/3}+0.58~\mathrm{fm}$ (except deuteron). Therefore
the dipole-nucleus cross-section into the positions of the
nucleons has been written [7] as
\begin{eqnarray}
\frac{d\sigma^{A}_{\mathrm{dip}}}{d^{2}b}&=&
2\int\prod_{i=1}^{A}\{d^{2}b_{i}T_{A}(b_{i})\}
\bigg{[}1-\prod_{i=1}^{A}S_{p}(\mathbf{r},\mathbf{b}-\mathbf{b}_{i};x)\bigg{]}\nonumber\\
&&{\approx}2\bigg{[}1-(1-\frac{T_{A}(\mathbf{b})}{2}\sigma^p_{\mathrm{dip}})^A\bigg{]}\nonumber\\
&&{\simeq}2\bigg{[}1-\exp(-AT_{A}(\mathbf{b})\sigma^p_{\mathrm{dip}}/{2})\bigg{]}.
\end{eqnarray}
 The nuclear $\gamma^*$-A interaction is defined as the $\gamma^*$-p interaction
 through
 \begin{eqnarray}
\sigma^{\gamma^{*}A}=\bigg{(}\frac{{\pi}R_{A}^2}{{\pi}R_{p}^2}\bigg{)}\sigma^{\gamma^{*}p},
\end{eqnarray}
with
\begin{eqnarray}
Q_{s,A}^{2}=\bigg{(}\frac{A{\pi}R_{p}^2}{{\pi}R_{A}^2}\bigg{)}^{1/\delta}Q_{s}^{2},
\end{eqnarray}
where for a nuclear target with the mass number $A$, the nuclear
radius is given by the usual parameterization\footnote{This
parametrization shows that the growth of the nuclear saturation
scale is faster than $A^{1/3}$ for large nuclei. }
$R_{A}=(1.12A^{1/3}-0.86A^{-1/3})~\mathrm{fm}$, and
$\delta=0.79{\pm}0.02$ [11]. The area of the proton is determined
to be  ${\pi}R_{p}^{2}=1.55{\pm}0.02~\mathrm{fm}^2$ [11].\\
The nuclear dipole cross section $\sigma^{A}_{\mathrm{dip}}$ is
dependent on the nuclear gluon distribution $g^{A}(x,Q^2)$ which
is defined in Ref.[38] with the replacement
$Q_{s}^{2}{\rightarrow}Q_{s,A}^{2}$,  by the following
form\footnote{The area of the nuclear target is replace by
$S^{A}=A^{2/3}S$, where $S$ is the nucleon target. }
\begin{eqnarray}
xg^A(x,Q^2)&=&f(A)\frac{3\sigma_{0}}{4\pi^2\alpha_{s}(Q^2)}\bigg{[}-Q^2\exp(-Q^2/Q^{2}_{s,A})\nonumber\\
&&+Q^{2}_{s,A}(1-\exp(-Q^2/Q^{2}_{s,A}))\bigg{]},
\end{eqnarray}
where the function $f(A)$ is defined to be $A^{2/3}$ [11] and
$Q^{2}_{s,A}$ is defined to be $Q_{s,A}^{2}=A^{1/3}Q_{s}^{2}$
[14]\footnote{ In the improved saturation model, a matching
between the dipole model gluon distribution and the collinear
approach is given in Ref.[39] using a leading order gluon
anomalous dimension $\gamma_{gg}$.}. Therefore, the ratio of the
color dipole cross section in the nuclear improved saturation
model, $\sigma^{A}_{\mathrm{dip}}/\sigma_{0}$, and at a given
impact parameter $b$ are given by
\begin{eqnarray}
\frac{\sigma^{A}_{\mathrm{dip}}}{\sigma_{0}}=1-\exp(-\frac{\pi^{2}r^{2}\alpha_{s}(\mu^{2})xg^A(\widetilde{x}_{f},\mu^{2})}{3\sigma_{0}}).
\end{eqnarray}
and
\begin{eqnarray}
\frac{d\sigma^{A}_{\mathrm{dip}}}{d^{2}b}=2\Big{[}1-
\exp\Big{(}-\frac{\pi^{2}r^{2}\alpha_{s}(\mu^{2})xg^{A}(\widetilde{x}_{f},\mu^{2})T_{A}(b)}{2N_{c}}\Big{)}
 \Big{]}.
\end{eqnarray}

\subsection{IV. Numerical Results}

In the leading order running coupling we set
$\Lambda_{\mathrm{QCD}}=120~\mathrm{MeV}$, which for the one-loop
coupling gives $\alpha_{s}(M_{Z}^2)=0.118$ and  other parameters
are defined by the following forms  according to Ref.[16] as
\begin{eqnarray}
\sigma_{0}=29.12~ \mathrm{mb},~ \lambda=0.277,~
x_{0}=0.41{\times}10^{-4},\nonumber\\
 m_{l}=0.14~\mathrm{GeV},~ m_{c}=1.40~\mathrm{GeV}.~~~~~~~~~~~~~~~~~
\end{eqnarray}
The results of our numerical studies of the saturation gluon
distribution in $eA$ processes, and comparison with  the nCETQ15
[40] for $\mathrm{Au}-197$ at $Q^2=16$ and $100~\mathrm{GeV}^2$
are shown in Fig.1.
\begin{figure}
\centerline{
\includegraphics[width=0.58\textwidth]{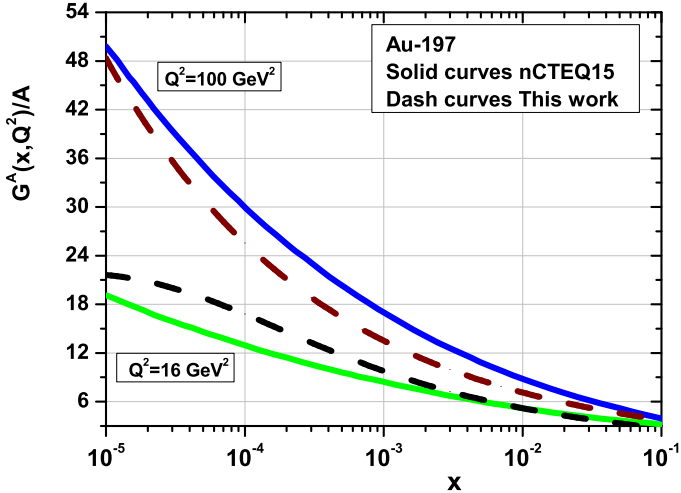}}
\caption{Results of the nuclear gluon distribution functions for
the nucleus of $\mathrm{Au}-197$. The gluon $G(x,Q^2)$
distributions per nucleon (dash curves) are shown as a function of
$x$ for $Q^2=16~\mathrm{GeV}^2$ and $Q^2=100~\mathrm{GeV}^2$,
respectively. For comparison, the solid curves show the results of
the nCTEQ15 [40] parametrization at the corresponding values of
$Q^2$, respectively.}\label{Fig2}
\end{figure}
\begin{figure}
\centerline{
\includegraphics[width=0.58\textwidth]{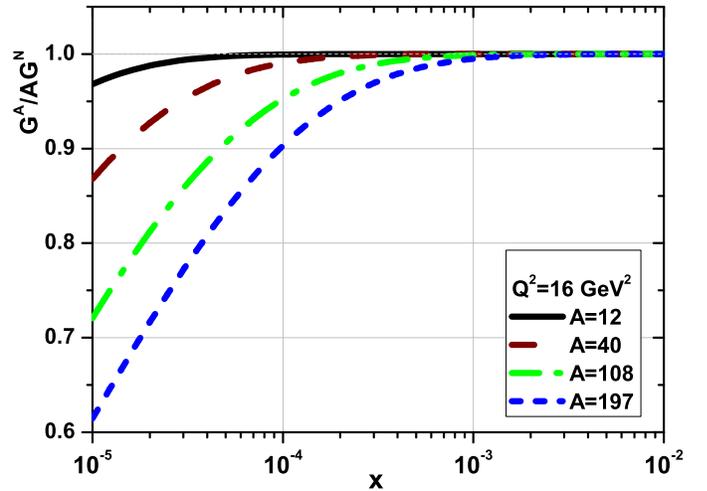}}
\caption{The ratios $\frac{1}{A}\frac{G^{A}(x,Q^2)}{G^{N}(x,Q^2)}$
of gluon distribution functions computed for different values of
$x$ for light and heavy nuclei including C-12, Ca-40, Ag-108,
Au-197, at $Q^2=16~\mathrm{GeV}^2$.}\label{Fig3}
\end{figure}
In this figure (i.e., Fig.1), we present results of the nuclear
gluon distribution function divided by $A$ for the heavy nucleus
of $\mathrm{Au}-197$ as a function of the momentum fraction $x$.
The dash curves show our results at $Q^2=16~\mathrm{GeV}^2$ and
$Q^2=100~\mathrm{GeV}^2$, respectively. They are compared to the
nCTEQ15 parametrization at the corresponding values of $Q^2$ given
by the solid curves, respectively. This figure indicates that the
results obtained from the present analysis are comparable with the
ones obtained from the nCTEQ15 parametrization.\\
The results for shadowing effects in the gluon distribution of
nuclei $\frac{1}{A}\frac{G^{A}(x,Q^2)}{G^{N}(x,Q^2)}$ at
$Q^2=16~\mathrm{GeV}^2$ for light and heavy nuclei including C-12,
Ca-40, Ag-108, Au-197 are shown in Fig.2. We observe that, as
expected, the shadowing effects are important for small
$x<10^{-3}$ and their magnitude decreases with a decrease of $x$
and with an increase of the atomic number A [41]. These results
are comparable with the results of Ref.[42] of gluon shadowing
correction\footnote{In Ref.[42], predictions for the gluon
shadowing correction from the $q\overline{q}G$ fluctuation of the
photon are shown by the following form
$\frac{1}{A}\frac{G^{A}(x,Q^2)}{G^{N}(x,Q^2)}{\sim}1-\frac{1}{A}\frac
{\Delta\sigma_{tot}(q\overline{q}G)}{\sigma_{tot}^{\gamma^{*}N}(x,Q^2)}$
, where $\Delta\sigma_{tot}(q\overline{q}G)$ is the inelastic
correction to the total cross section
$\sigma_{tot}^{\gamma^{*}N}(x,Q^2)$.} corresponding to the
$|q\overline{q}G>$ Fock component of the photon containing one
gluon. These behaviors are observable in other phenomenological
parametrizations, such as GBW,KST [43]
,BGBK and IP-sat models.\\
\begin{figure}
\centerline{
\includegraphics[width=0.58\textwidth]{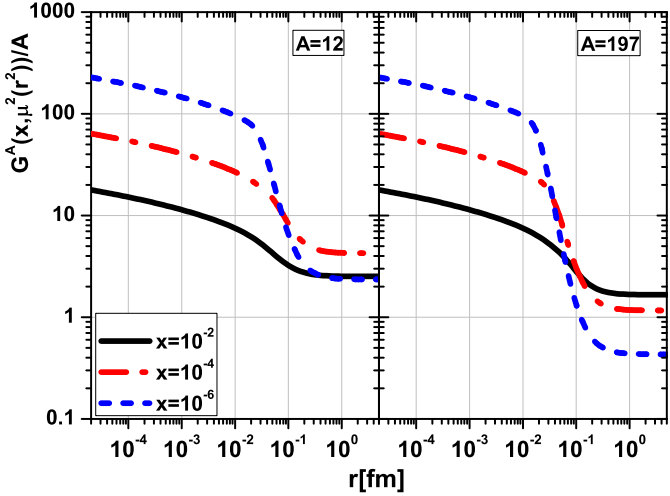}}
\caption{Results of the nuclear gluon distribution functions for
the nucleus of $\mathrm{C}-12$ and $\mathrm{Au}-197$. The gluon
$G^A(x,\mu^2)$ distributions per nucleon are shown as a function
of the dipole transverse size, $r$, for $x=10^{-2}$ (solid),
$x=10^{-4}$ (dashed-dot) and $x=10^{-6}$ (short dashed),
respectively.}\label{Fig4}
\end{figure}
In the improved saturation model, the connection between  the
nuclear dipole cross section, $\sigma^{A}_{dip}$, and the
integrated nuclear gluon density is crucial for describe the
exclusive processes in  eA collisions [4]. The evolution of the
analytical nuclear gluon distribution divided by A for A=12 and
197 as a function of the dipole transverse size, $r$, is shown in
Fig.3. In this figure (i.e., Fig.3), we observe a slow decrease of
the nuclear gluon distribution in the large dipole domain, for
$x=10^{-2}$ for light and heavy nuclei. This behavior in the large
dipole domain is strongly decreases as the Bjorken value decrease
and the number of nucleons in a nuclear target increase.\\
Figure 4 quantifies the size of the dipole cross sections as a
function of the mass number A. It presents the ratio
$\sigma^{A}_{\mathrm{dip}}/\sigma_{0}$ as a function of $r$ for
light and heavy nuclei including C-12, Ca-40, Ag-108, Au-197 and
the free proton. The ratio for the free proton (short dashed-dot)
is compared with the GBW model (short dot-thin, Eq.(2)) in a wide
range of $r$ for $x=10^{-3}$. It is clearly seen where saturation
is visible for the free proton at $r{\sim}1~\mathrm{fm}$ and this
value decrease as $A$ increases.
\begin{figure}
\centerline{
\includegraphics[width=0.58\textwidth]{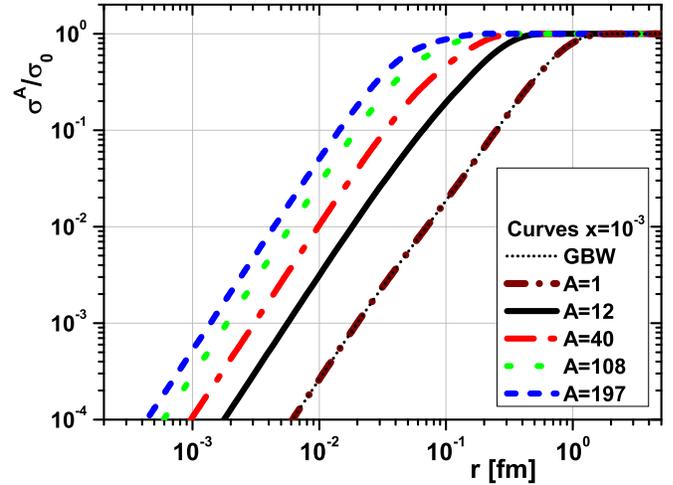}}
\caption{The ratio $\sigma^{A}_{\mathrm{dip}}/\sigma_{0}$ as a
function of $r$ at $x=10^{-3}$ for light and heavy nuclei
including C-12 (solid), Ca-40 (dashed-dot), Ag-108 (dot), Au-197
(short-dash) and the free proton (short dashed-dot). The ratio for
the free proton (short dashed-dot)  is compared with the GBW model
(short dot-thin).}\label{Fig5}
\end{figure}
The improved saturation model in nuclei gives a similar behavior
of the ratio $\sigma^{A}_{\mathrm{dip}}/\sigma_{0}$ in comparison
with the GBW saturation model at low $x$ in a wide range of the
dipole transverse size $r$. Calculations have been performed at
the Bjorken variable $x$ to vary in the interval
$x=10^{-6}...10^{-2}$ for Au-197 in Fig.5. The improved saturation
model for nuclei gives a good description of the ratio
$\sigma^{A}_{\mathrm{dip}}/\sigma_{0}$ in comparison with the GBW
saturation model at low $x$ in a wide range of the momentum
transfer $Q^{2}$. In Fig.5 we observe that, in the interval
$2.10^{-2}~\mathrm{fm}{\lesssim}r{\lesssim}5.10^{-1}~\mathrm{fm}$,
a depletion occurs  for $x<10^{-3}$. This depletion is strongly
dependence to the mass number A. In Fig.6 this behavior for the
light and heavy nuclei is shown for $x=10^{-6}$, which
significantly enhances the importance of the nonlinear corrections
for heavy nuclei compared to the proton case.
\begin{figure}
\centerline{
\includegraphics[width=0.58\textwidth]{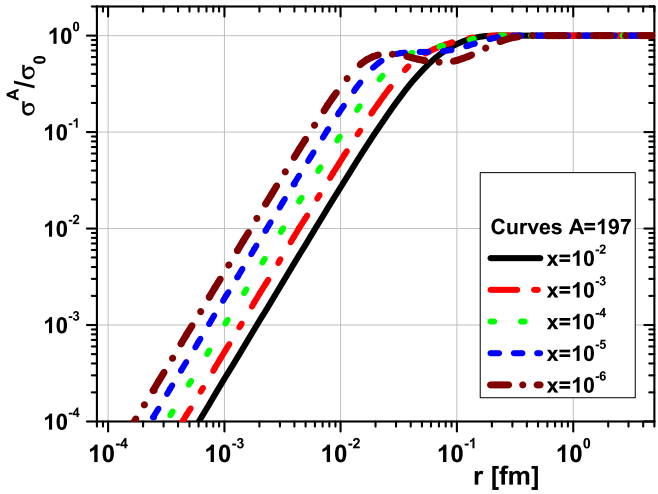}}
\caption{The extracted ratio
$\sigma^{A}_{\mathrm{dip}}/\sigma_{0}$ as a function of $r$ at
$x=10^{-6}...10^{-2}$ (curves from left to right, respectively)
for Au-197.}\label{Fig6}
\end{figure}
This effect is visible in the range
$1.75~\mathrm{GeV}^2<\mu^2<3.3~\mathrm{GeV}^2$ at very low $x$
(i.e., $x=10^{-6}$) for heavy nuclei. One can see from the figure
6 that the nonlinear effects clearly become more important with
increasing A, for small values of $x$ and $Q^{2}$. Indeed , the
deviation from unity in this ratio is an indication of color
transparency. A depletion in this ratio is called $"$shadowing$"$,
whereas an enhancement is called $"$anti-shadowing$"$. The
anti-shadowing  is related to the coherent multiple scattering
where it introduces the medium size enhanced (in powers of
$A^{1/3}$) nuclear effects [44-49]. The nuclear shadowing is
controlled by the interplay of photon lifetime and coherent time
fluctuations for transition between no
shadowing and saturated shadowing at very small $x$ [50,51].\\
Depilation and enhancement observed for heavy nuclei in Figs.5 and
6 at very small $x$ at the interval $10^{-2}<r<10^{0}~\mathrm{fm}$
can be dependent on the important of the gluon shadowing from
higher $|q\overline{q}G>$ Fock component of the photon in dipole
model, which leads to the renormalization of the dipole cross
section\footnote{For further discussion please see Ref.[42].}.\\
In Fig.7, we have plotted the ratio
$\sigma^{2A}_{\mathrm{dip}}/\sigma^{2}_{0}$ (in the following
$\sigma_0^2$ and $\sigma^{2A}$ are suggested to be
$\sigma_0^{p+n}$ and $\sigma_A^{p+n}$) for the
diffractive\footnote{The cross section for the diffractive
$q\overline{q}$ production reads
$$
\frac{d\sigma_{L,T}^{D}}{dt}|_{t=0}=\int dz d^{2}\mathbf{r}
|\Psi_{L,T}(\mathbf{r},z,Q^{2})|^{2}\sigma^{2}_{\mathrm{dip}}(\widetilde{x}_{f},\mathbf{r}),
$$
where $t=\Delta^2$, and $\Delta$ is the four-momentum transferred
into the diffractive system from the proton.} $q\overline{q}$
production in the color singlet state as a function of $r$ at
$x=10^{-6}$ for light and heavy nuclei including C-12,
 Ag-108, Pb-208 and the free proton. The
diffractive $\gamma^{*}A{\rightarrow}q\overline{q}A'$ cross
section is proportional to $\sigma^{2A}_{\mathrm{dip}}(x,r)$,
where at small values of the diffractive mass $M^{2}\sim Q^{2}$
the elastic scattering of the $q\overline{q}$ pair dominates. In
this figure (i.e., Fig.7), we observe that the saddle point
increases as the mass number A increases as the magnitude of
shadowing is increased using the nuclear density function within
the higher Fock component of the photon containing gluons. This
behavior of the ratio $\sigma^{2A}_{\mathrm{dip}}/\sigma^{2}_{0}$
for heavy nuclei is deeper than the ratio
$\sigma^{A}_{\mathrm{dip}}/\sigma_{0}$.\\
\begin{figure}
\centerline{
\includegraphics[width=0.58\textwidth]{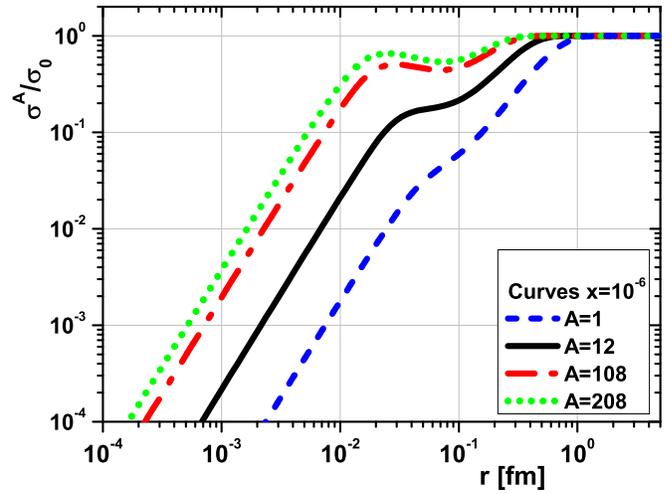}}
\caption{Results of the nonlinear effects due to the mass number A
for the ratio $\sigma^{A}_{\mathrm{dip}}/\sigma_{0}$ as a function
of $r$ at $x=10^{-6}$ or a wide range of nuclei including C-12,
 Ag-108, Pb-208 and the free proton.}\label{Fig7}
\end{figure}
\begin{figure}
\centerline{
\includegraphics[width=0.58\textwidth]{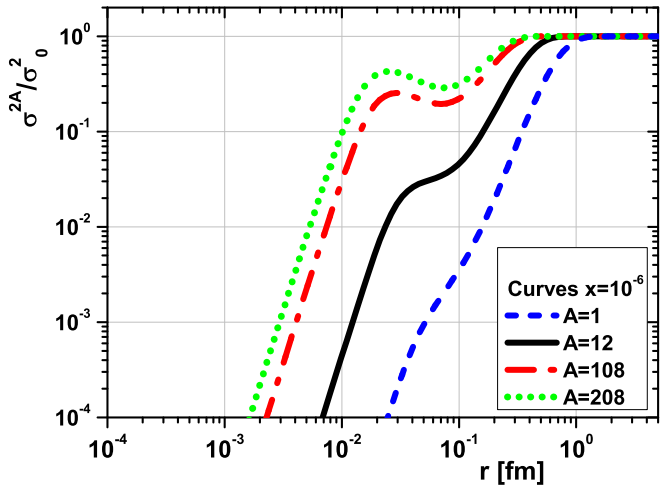}}
\caption{Results of the nonlinear effects due to the mass number A
in the the simplest case of the $q\overline{q}$ system  for the
ratio $\sigma^{2A}_{\mathrm{dip}}/\sigma^{2}_{0}$ as a function of
$r$ at $x=10^{-6}$ for light and heavy nuclei including C-12,
 Ag-108, Pb-208 and the free proton.}\label{Fig8}
\end{figure}
\begin{figure}
\centerline{
\includegraphics[width=0.58\textwidth]{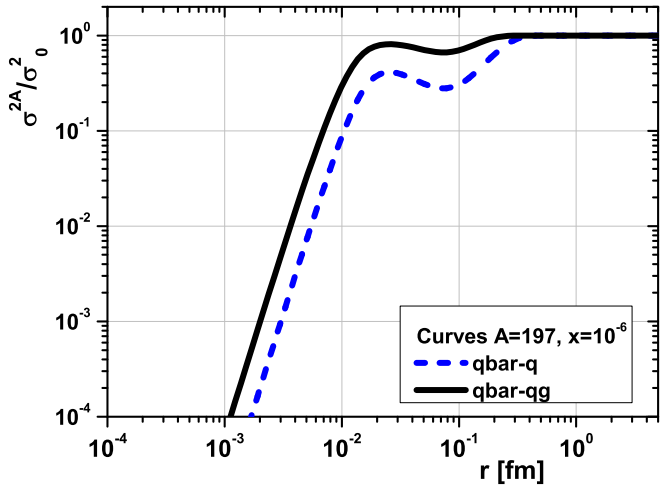}}
\caption{ Comparing between the $q\overline{q}$ and
$q\overline{q}g$ components of the diffractive system in the ratio
$\sigma^{2A}_{\mathrm{dip}}/\sigma^{2}_{0}$ as a function of $r$
at $x=10^{-6}$ for Au-197.}\label{Fig9}
\end{figure}
In Fig.8, we have added the $q\overline{q}g$ contribution (due to
gluon production in the final diffractive state) for the
diffractive processes at larger values of the mass
$M^{2}{\gg}Q^{2}$ by a weight factor
$C_{A}/C_{F}=2N_{C}^{2}/(N_{C}^{2}-1)$ with $C_{A}=N_{c}=3$ and
$C_{F}=\frac{N_{C}^{2}-1}{N_{C}}=\frac{4}{3}$ where $N_{C}$ is the
number of colors [17,26-27]. This component was computed in the
two gluons exchange approximation with  a color octet dipole
$8\overline{8}$ where the coupling of two $t$-channel gluons is
relative by the weight factor\footnote{ The color dipole cross
section for exchange of a two gluon system for octet dipole reads
$$
\sigma^{p}_{\mathrm{dip}}=\sigma_{0}\{1-\exp(-\frac{C_{A}}{C_{F}}\frac{\pi^{2}r^{2}\alpha_{s}(\mu^{2})xg(\widetilde{x}_{f},\mu^{2})}{3\sigma_{0}})\}
$$
}, as the saddle point behavior in the region
$10^{-2}<r<10^{0}~\mathrm{fm}$ decreases at $x=10^{-6}$ for
Au-197. This behavior is completely tamed if we consider the
propagation of the $|q\overline{q}>$ Fock component of the photon
in a nuclear medium by higher order of shadowing corrections as
$|q\overline{q}>=|q\overline{q}>+|q\overline{q}G>+|q\overline{q}2G>+...$
due to different Fock states [42].\\
\begin{figure}
\centerline{
\includegraphics[width=0.58\textwidth]{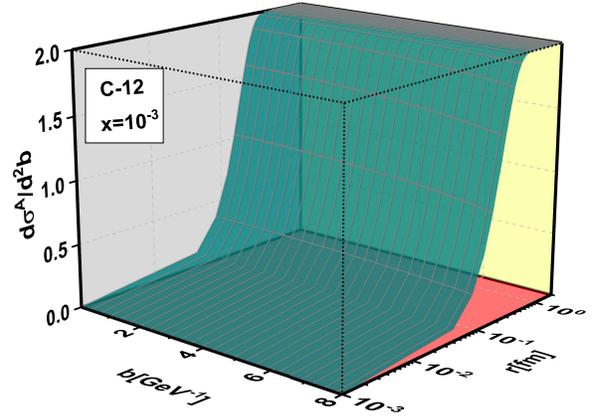}}
\caption{ The nuclear dipole cross section at impact parameter $b$
as a function of $r$ and $b$ at $x=10^{-3}$ for
C-12.}\label{Fig10}
\end{figure}
\begin{figure}
\centerline{
\includegraphics[width=0.58\textwidth]{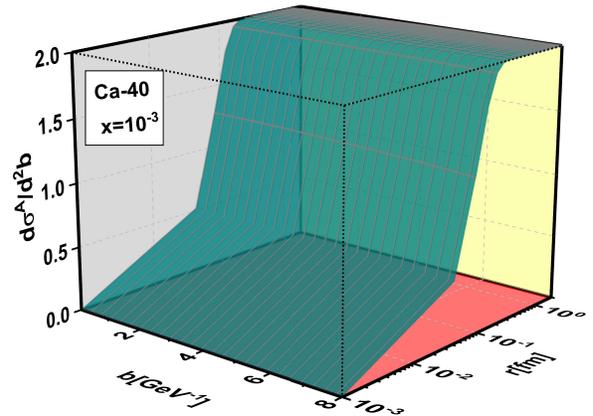}}
\caption{ The same as Fig.9 for Ca-40.}\label{Fig11}
\end{figure}
In Figs. 9 and 10, we consider the differential cross section
$d\sigma_{\mathrm{dip}}^{A}/d^{2}b$ at a given impact parameter
$b$, using the definition of the total cross section of the
$q\overline{q}$ pair on the proton $\sigma^{p}_{q\overline{q}}$,
by the integrated Woods-Saxon distribution $T_{A}(b)$ scaled by
the number of nucleons, for $x=10^{-3}$ [38,52]. In these figures
(i.e., Figs.9 and 10), the nuclear dipole scatters at impact
parameter $b$ are calculated for the nuclei C-12 and Ca-40 in a
wide range of the parameters $b$ and $r$, respectively. We observe
that the saturation is visible at $r{\simeq}1~\mathrm{fm}$ for
C-12 in a wide range of $b$, $0{\leq}b{\leq}8~\mathrm{GeV^{-1}}$,
and increase towards lower $r$ (i.e., $r<1~\mathrm{fm}$) when the
mass number A increases (see Fig.10 for Ca-40). These 3D figures
have a broken line in the behavior of
$d\sigma_{\mathrm{dip}}^{A}/d^{2}b$ as it increases from
approximately $\simeq$0.1 to 0.5 with an increase $A$ from 12 to
40, respectively. We see that the two functions for C-12 and Ca-40
differ in the small-$r$ region where the running of the gluon
distribution starts to play a significant role, with an increase
of the mass number $A$. Indeed, the behavior of the
$d\sigma_{\mathrm{dip}}^{A}/d^{2}b$ is directly dependent on the
gluon density and the mass number $A$. These behaviors clearly
indicate that the IP saturation model can be used to study nuclear
effects in the
future experiments at electron-ion colliders.\\

\subsection{V. Conclusions}

In this paper, we studied the improved saturation model for nuclei
with respect to the gluon density obtained within the color dipole
approach. The nuclear cross-section is evaluated by implying the
impact of the nuclear gluon density at small $x$. We presented the
study of the shadowing in deep-inelastic scattering off nuclei in
the kinematic regions accessible by future electron-ion colliders.
The dipole cross sections are considered in the description of the
inclusive and diffractive DIS at small $x$ in a wide range of the
mass number $A$. The ratio $\sigma^{A}_{\mathrm{dip}}/\sigma_{0}$
due to the nuclear effects is similar with the GBW saturation
model at low $x$, although the saturation region expands with the
increase of the mass number $A$. A saddle-shaped behavior is
predicted at very low $x$ for heavy nuclei in a range
$2{\times}10^{-2}{\lesssim}~r~{\lesssim}2{\times}10^{-1}~\mathrm{fm}$
due to the nonlinear effects. In the diffractive DIS processes
where the component $q\overline{q}g$ deviates from the GBW and CGC
models, the behavior at very low $x$ for heavy nuclei is tamed.
This behavior increases the saturation region with the increase of
the mass number of $A$.\\
Nuclear corrections to the impact parameter dependent dipole cross
section in a wide range of the impact parameter $b$ and the dipole
size $r$ are considered. The saturation region in the IP-Sat model
increases as $r$ decreases and the mass number of $A$ increases,
in a wide range of $b$. Indeed, we have tested the IP-Sat model
with impact parameter dependence with increases of the mass number
of $A$. While the influence of the impact parameter structure
decreases as the mass number of $A$ increases and gives a
possibility to test various models for the nuclear dipole cross
section at small $x$ at future colliders such as EIC and the
LHeC.\\


\subsection{ACKNOWLEDGMENTS}
The author is grateful to Razi University for the financial
 support of this project.\\



\section{References}

1. LHeC Collaboration and FCC-he Study Group, P. Agostini
et al., J. Phys. G: Nucl. Part. Phys. {\bf48}, 110501(2021).\\
2. A.Accardi et al., Eur.Phys.J.A {\bf52}, 268 (2016).\\
3. R. Abdul Khalek et al., (2021), arXiv:2103.05419
[physics.ins-det].\\
4. D.A Fagundes and M.V.T.Machado, Phys.Rev. D {\bf107}, 014004
(2023).\\
5. V.P.Goncalves and M.V.T.Machado, Eur.Phys.J.C {\bf37},299
(2004).\\
6. M.A.Betemps
 and M.V.T.Machado, arXiv:0906.5593.\\
7. C.Marquet, Manoel R.Moldes and P.Zurita, Phys.Lett.B {\bf772},
607 (2017).\\
8. N.Armesto, Eur.Phys.J.C {\bf26}, 35 (2002).\\
9. Ya-Ping Xie and Victor P. Goncalves, Phys.Rev. D {\bf105},
014033 (2022).\\
10. A. M.Stasto, K.Golec-Biernat and J.Kwiecinski, Phys.Rev.Lett.
{\bf86}, 596 (2001).\\
11. Nestor Armesto, Carlos A. Salgado, Urs Achim Wiedemann,
Phys.Rev.Lett. {\bf94}, 022002 (2005).\\
12. J. Raufeisen, Acta
Phys.Polon. B {\bf36}, 235 (2005).\\
13. N.N.Nikolaev, W.Schafer, B.G.Zakharov and V.R.Zoller, JETP
Letters {\bf84}, 537 (2006).\\
14. F.Carvalho, F.O.Duraes, F.S.Navarra and S.Szpigel, Phys.Rev.C
{\bf79}, 035211 (2009).\\
15. K. Golec-Biernat and S.Sapeta, JHEP {\bf03},
102 (2018).\\
16. K.Golec-Biernat, J.Phys.G {\bf28}, 1057 (2002).\\
17. J.Rausch, V. Guzey and M. Klasen, Phys.Rev.D {\bf107}, 054003
(2023).\\
18. Y.Hatta, Nucl.Phys.A {\bf00}, 1 (2020).\\
19. O.Bruning, A. Seryi
and S. Verdu-Andres, Front.in Phys. {\bf10}, 886473 (2022).\\
20. M.Klasen, K.Kovarik and J.Potthoff, Phys. Rev. D {\bf95}, 094013 (2017).\\
21. J.D.Bjorken, J.B.Kogut and D.E.Sopper, Phys.Rev.D {\bf3}, 1382
(1971).\\
22. N.N.Nikolaev and B.G.Zakharov, Z.Phys.C {\bf49}, 607 (1991);
Z.Phys.C {\bf53}, 331 (1992); Z.Phys.C {\bf64}, 651 (1994); JETP
{\bf78}, 598 (1994).\\
23.  A.H.Mueller, Nucl.Phys.B {\bf415}, 373 (1994); A.H.Mueller
and B.Patel, Nucl.Phys.B {\bf425}, 471 (1994); A.H.Mueller,
Nucl.Phys.B {\bf437}, 107 (1995).\\
24. J.Bartels, K.Golec-Biernat and H.Kowalski, Phys. Rev.
D{\bf66}, 014001 (2002).\\
25. H.Kowalski and D.Teaney, Phys. Rev.
D{\bf68}, 114005 (2003).\\
26. K.Golec-Biernat, Acta.Phys.Polon.B{\bf33}, 2771 (2002);
Acta.Phys.Polon.B{\bf35}, 3103 (2004); J.Bartels, K.Golec-Biernat
and H.Kowalski, Acta.Phys.Polon.B{\bf33}, 2853 (2002); E.Iancu,
K.Itakura and S.Munier, Phys.Lett.B {\bf590}, 199 (2004);
J.R.Forshaw and G.Shaw, JHEP {\bf12}, 052 (2004).\\
27. K.Golec-Biernat, M.W$\ddot{u}$sthoff, Phys.Rev.D 59, 014017
(1998); Phys.Rev.D 60, 114023 (1999).\\
28. A.H. Mueller, Nucl.Phys.B {\bf335}, 115 (1990).\\
29. L. McLerran and R. Venugopalan, Phys. Rev. D {\bf49}, 2233
(1994).\\
30. I. Balitsky, Nucl. Phys. B{\bf463}, 99 (1996).\\
31. Y. V. Kovchegov,
Phys. Rev. D{\bf60},  034008(1999).\\
32. Y. V. Kovchegov, Phys. Rev. D{\bf61}, 074018 (2000).\\
33. A.M.Stasto, K.Golec-Biernat and J.Kwiecinski,
Phys.Rev.Lett.{\bf86}, 596 (2001).\\
34. E. Iancu and R. Venugopalan, Quark-Gluon Plasma 3, (2004) 249, World Scientific Publishing Co Pte Ltd.\\
35. G.Watt and H.Kowalski, Phys. Rev. D{\bf78}, 014016 (2008).\\
36. Yuan-Yuan Zhang and Xin-Nian Wang, Phys.Rev.D {\bf105}, 034015 (2022).\\
37. A. Capella, A. Kaidalov, C. Merino, D. Pertermann and J. Tran
Thanh Van,     Eur.Phys.J.C {\bf5}, 111 (1998).\\
38. R.S.Thorne, Phys.Rev.D {\bf71}, 054024 (2005).\\
39. R. D. Ball and S. Forte, Phys. Lett. B {\bf335}, 77 (1994).\\
40. K.Kovarik, A.Kusina, T.Jezo, et al., Phys.Rev.D {\bf93},
085037 (2016).\\
41. F.Muhammadi and B.Rezaei, Phys.Rev.C {\bf106}, 025203
(2022).\\
42. M.Krelina and J.Nemchik, Eur. Phys. J. Plus {\bf135}, 444
(2020).\\
43. B.Z.Kopeliovich, A.Schafer and A.V.Tarasov, Phys. Rev. D
{\bf62},
054022 (2000).\\
44. G.R.Boroun, B.Rezaei and S.Heidari, Int.J.Mod.Phys.A {\bf32},
1750197 (2017).\\
45. G.R.Boroun and B.Rezaei, Phys.Rev.C {\bf107}, 025209 (2023).\\
46. S.Heidari, B.Rezaei and G.R.Boroun, Int.J.Mod.Phys.E {\bf26},
1750067 (2017).\\
47. E.R.Cazaroto, F.Carvalho, V.P.Goncalves and F.S.Navarra,
Phys.Lett.B {\bf669}
, 331(2008).\\
48. X.Guo and J.Li, Nucl. Phys. A {\bf783}, 587 (2007).\\
49. K. Golec-Biernat et al., Nucl. Phys. B {\bf527}, 289 (1998).\\
50. G.R.Boroun, Eur.Phys.J.C  {\bf82}, 740 (2022).\\
51. G.R.Boroun, Eur.Phys.J.C {\bf83}, 42 (2023).\\
52. H.Kowalski and D.Teaney, Phys.Rev.D {\bf68}, 114005 (2003).\\


\end{document}